\documentclass[preprint,showpacs,preprintnumbers,amsmath,amssymb,nofootinbib]{revtex4}
\usepackage{bbm}
\usepackage{amsfonts}
\usepackage{slashed}
\usepackage{subfigure}
\usepackage{array}

\usepackage{booktabs}
\usepackage{color, soul}
\usepackage{mathrsfs}
\usepackage{epsfig}
\usepackage{CJK}
\usepackage{graphicx}
\usepackage{dcolumn}
\usepackage{bm}
\usepackage{amsmath}
\usepackage{amssymb}

\newcommand{\p}{\partial}

\newcommand{\sh}{{\!\slash}}
\newcommand{\met}{\ \sh{\hspace{-0.3cm}$E_T$}\ }

\let\jnfont=\rm
\def\NPB#1,{{\jnfont Nucl.\ Phys.\ B }{\bf #1},}
\def\PLB#1,{{\jnfont Phys.\ Lett.\ B }{\bf #1},}
\def\EPJC#1,{{\jnfont Eur.\ Phys.\ Jour.\ C }{\bf #1},}
\def\PRD#1,{{\jnfont Phys.\ Rev.\ D }{\bf #1},}
\def\PRL#1,{{\jnfont Phys.\ Rev.\ Lett.\ }{\bf #1},}
\def\MPLA#1,{{\jnfont Mod.\ Phys.\ Lett.\ A }{\bf #1},}
\def\JPG#1,{{\jnfont J.\ Phys.\ G}{\bf #1},}
\def\CTP#1,{{\jnfont Commun.\ Theor.\ Phys.\ }{\bf #1},}
\def\ZPC#1,{{\jnfont Z.\ Phys.\ C }{\bf #1},}
\def\JHEP#1,{{\jnfont JHEP \ }{\bf #1},}
\def\Rv{\not{\hbox{\kern-1pt $R$}}}
\def\p{\not{\hbox{\kern-3pt $p$}}}

\begin{document}


\title{A light SUSY dark matter after CDMS-II, LUX and LHC Higgs data}

\author{Junjie Cao$^{1,2}$, Chengcheng Han$^3$, Lei Wu$^4$, Peiwen Wu$^3$, Jin Min Yang$^3$}

\affiliation{ $^1$  Department of Physics,
        Henan Normal University, Xinxiang 453007, China \\
  $^2$ Center for High Energy Physics, Peking University,
       Beijing 100871, China \\
  $^3$ State Key Laboratory of Theoretical Physics,
      Institute of Theoretical Physics, Academia Sinica, Beijing 100190,
      China\\
      $^4$ ARC Center of Excellence for Particle Physics at the Terascale, School of Physics,
       The University of Sydney, NSW 2006, Australia  }


\begin{abstract}
In SUSY, a light dark matter is usually accompanied by light scalars to achieve the correct relic density, which opens
new decay channels of the SM-like Higgs boson. Under current experimental constraints including the latest LHC Higgs data
and the dark matter relic density, we examine the status of a light neutralino
dark matter in the framework of NMSSM and confront it
with the direct detection results of CoGeNT, CDMS-II and LUX. We have the
following observations: (i) A dark matter as light as 8 GeV is still
allowed and its scattering cross section off the nucleon can be large enough
to explain the CoGeNT/CDMS-II favored region;
(ii) The LUX data can exclude a sizable part of the allowed parameter space, but
still leaves a light dark matter viable;
(iii) The SM-like Higgs boson can decay into the light dark matter pair with an invisible branching ratio reaching $30\%$ under the current LHC
Higgs data, which may be tested at the 14 TeV LHC experiment.
\end{abstract}
\maketitle


\section{\label{sec:level1}Introduction}

As one of the most compelling evidences for new physics beyond the Standard Model (SM),
the cosmic dark matter (DM) has been widely studied in particle physics \cite{DM_before_1,DM_before_2,DM_before_3,DM_before_4,DM_before_5,DM_before_6,DM_before_7}.
Recently, the CDMS-II collaboration observed three events which can be explained
by a light DM with mass about 8.6 GeV and a spin-independent DM-nucleon scattering cross
section of about $1.9 \times 10^{-5}$ pb \cite{CDMS2}.
The existence of such a light DM seems to be corroborated by other direct detections such as
the CoGeNT \cite{CoGeNT1,CoGeNT2}, CRESST \cite{CRESST} and DAMA/LIBRA \cite{DAMA}. Moreover,
a light DM is also hinted by Fermi-LAT, a satellite-based DM indirect detection experiment \cite{Fermi_LAT_experiment}.
Recent analysis of the Fermi-LAT data exhibits peaks in the gamma-ray spectrum at energies
around 1-10 GeV, which could be interpreted in terms of the annihilation of a DM with mass
low than about 60 GeV into leptons or bottom quarks \cite{Fermi_LAT_analysis_1,Fermi_LAT_analysis_2,Fermi_LAT_analysis_3,Fermi_LAT_analysis_4,Fermi_LAT_analysis_5}.
About these experimental results, it should be noted that they are not completely consistent with each other, and
more seriously, they conflict with the XENON data \cite{XENON100} and
the latest LUX result \cite{lux}.
So the issue of light DM leaves unresolved and will be a focal point both
experimentally and theoretically.
On the experimental side, many experiments like LUX, XENON, CDMS and CDEX \cite{CDEX_1,CDEX_2} will
continue their searches, while on the theoretical side we need to examine
if such a light dark matter can naturally be predicted in popular new physics
theories such as low energy supersymmetry (SUSY).

Previous studies \cite{cao_cdms,dark-higgs} showed that, in the framework of the
Next-to-Minimal Supersymmetric Standard Model (NMSSM) \cite{nmssm_review}, a light
neutralino DM around 10 GeV is allowed by the collider constraints and DM relic density
(in contrast such a light DM is not easy to obtain in
the MSSM \cite{MSSM_DM_han} or CMSSM \cite{MSSM-light-DM-sbottom}).
In NMSSM, due to the presence of a singlet superfield $\hat{S}$,
we have five neutralinos, three CP-even Higgs bosons ($h_{1,2,3}$) and two CP-odd Higgs
bosons ($a_{1,2}$) \cite{nmssm_review}. The mass eigenstates of neutralinos are the
mixture of the neutral singlino ($\tilde{S}$), bino ($\tilde{B}$), wino ($\tilde{W}^0$)
and higgsinos ($\tilde{H}^0_u$, $\tilde{H}^0_d$); while the CP-even (odd) Higgs mass
eigenstates are the mixture of the real (imaginary) part of the singlet scalar $S$
and the CP-even (odd) MSSM doublet Higgs fields. An important feature of the NMSSM is that the lightest CP-even (odd) Higgs boson $h_1(a_1)$ can be singlet-like and very light, and the lightest neutralino
($\tilde\chi^0_1$) can be singlino-like and also very light. As a result,
the spin-independent neutralino-nucleon scattering cross section can be enhanced to
reach the CDMS-II value by the $t-$channel mediation of a light $h_1$ \cite{cao_cdms,dark-higgs}.
Meanwhile, the DM relic density can be consistent with the measured value
either through the $s-$channel resonance effect of $h_1(a_1)$ in DM annihilation or through the annihilation
into a pair of light $h_1$ or $a_1$ \cite{cao_cdms,dark-higgs}.

Note that such a light DM in the NMSSM should be re-examined because
the latest LHC data may give severe constraints. Due to the presence of
a light DM and concurrently a light  $a_1$ or $h_1$,
the SM-like Higgs boson ($h_{SM}$) can have new decays
$h_{SM} \to \tilde\chi^0_1 \tilde\chi^0_1$
and $h_{SM} \to h_1 h_1(~{\rm or}~ a_1a_1)$ \cite{cao_light_higgs}.
As analyzed in \cite{belanger_fit}, such decays may be subject to
stringent constraints from the current LHC Higgs data \cite{lhc_data}.
Besides, since a certain amount of higgsino component in $\tilde\chi^0_1$ is
needed to strengthen the coupling of $h_1\tilde\chi^0_1\tilde\chi^0_1$
(or $a_1\tilde\chi^0_1\tilde\chi^0_1$) which is necessary for the DM annihilation,
the higgsino-dominated neutralinos and the chargino $\tilde\chi^+_1$
are generally not very heavy and will be constrained by the searches
for events with three leptons and missing transverse momentum ($3\ell$+\met) at 8 TeV LHC \cite{atlas_cms_ew,Ellwanger_higgsino_singlino_sector}.
In this work, we consider these latest LHC data and examine the status of
a light DM in the NMSSM.

We note that a recent study \cite{french_dm} tried to explain the CDMS-II results in terms of
a light DM in the NMSSM. Compared to \cite{french_dm} which only studied three representative benchmark points, we perform a numerical scan under various experimental constraints and display the allowed parameter space in comparison with the the direct detection results of CoGeNT, CDMS-II and LUX. We also perform a global fit of the Higgs data using the package \textsf{HiggsSignals-1.0.0} \cite{higgssignal}, in which we further consider the latest LHC results of Higgs invisible decay from the channel $pp\to ZH$ \cite{zh_8}. Moreover, we consider the constraints from the searches for events with $3\ell$+\met signal at 8 TeV LHC \cite{atlas_cms_ew,Ellwanger_higgsino_singlino_sector}.

The paper is organized as follows.
In Sec.II we list the experimental constraints and describe our scan.
In Sect.III we present our results and perform detailed analysis.
Finally, we draw our conclusions in Sec.IV.

\section{\label{Scanning} A numerical scan}
In order to reduce the number of free parameters in our scan over the NMSSM parameter space, we make some assumptions on the parameters that do not influence
DM properties significantly. Explicitly speaking, we fix gluino mass and all the soft mass parameters in
squark sector at 2 TeV, and those in slepton sector at 300 GeV. We also assume the soft trilinear couplings
$A_t=A_b$ and let them vary to tune the Higgs mass. Moreover, in order to predict a bino-like light DM
and also to avoid the constraints from $Z$ invisible decay \cite{zdecay},
we abandon the GUT relation between $M_1$ and $M_2$. The free
parameters are then $\tan\beta, \lambda, \kappa, A_\lambda, A_\kappa$ in the Higgs sector,
the gaugino and higgsino mass parameters $M_1, M_2$ and $\mu$, and the soft trilinear couplings of the
third generation squarks $A_t$. In this work, we define all these parameters at $2 \,{\rm TeV}$ scale and
adopt the Markov Chain Monte Carlo (MCMC) method to scan the following parameter
ranges using $\textsf{NMSSMTools-4.0.0}$ \cite{nmssmtools}:
\begin{eqnarray}
&& 1 < \tan\beta < 40,~ 0 < \lambda < 0.7, ~ 0 < |\kappa| < 0.7, \nonumber\\
&& 0 < |A_\kappa| < 2 ~{\rm TeV}, ~ 0 < A_\lambda < 5 ~{\rm TeV}, ~ |A_t| < 5 ~{\rm TeV}, \nonumber\\
&& 0 < |M_1| < 0.6 ~{\rm TeV}, ~ 0.32 ~{\rm TeV} < M_2 < 0.6 ~{\rm TeV},~ 0.1 ~{\rm TeV}< \mu < 0.6 ~{\rm TeV}.
\end{eqnarray}
Note here that the ranges of $\lambda$ and $\kappa$ are motivated to avoid Landau pole, generally corresponding to the requirement of $\sqrt{\lambda^2+\kappa^2}\lesssim 0.7$. This has been encoded in $\textsf{NMSSMTools-4.0.0}$ including the consideration of the interplay between $\lambda$ and $\kappa$ in the renormalization group running. A relatively small $\mu$ is chosen to avoid strong cancelation in getting the $Z$ boson mass \cite{nmssm_review}, and as we will see below, the upper bound of $600\, {\rm GeV}$ for $\mu$ here suffices our study and does not affect our main conclusions. Also note that we artificially impose a lower bound of 320 GeV for $M_2$. This is motivated by the fact that $M_2$ in our study is not an important parameter, and that as required by the $3\ell$+ \met constraint $M_2$ should be larger than about 320 GeV in the simplified model discussed in \cite{atlas_cms_ew} (also see the constraint (viii) discussed below). The relevant $\chi^2$ function for the MCMC scan is build to guarantee the DM relic density and the SM-like Higgs boson mass around their measured values.
In our discussion, we consider the samples surviving the following constraints:
\begin{itemize}
\item[(i)] $123 \, {\rm GeV} \leq m_{h_{SM}} \leq 127 \, {\rm GeV}$ and $m_{\tilde{\chi}_1^0} \leq m_{h_{SM}}/2$.

\item[(ii)]  {\em The constraints from B-physics}.
The light CP-even/odd Higgs bosons can significantly affect the B-physics observables.
Especially, the precise measurements of
radiative decays $\Upsilon \to h_1 \gamma, a_1 \gamma$ \cite{upsilon}, $B \to X_s \gamma $ \cite{bsg}
and $B_s \to \mu^+\mu^-$ \cite{bsmm} can give stringent constraints.
So we require the samples to satisfy these B-physics bounds at 2$\sigma$ level.

\item[(iii)] {\em DM relic density}.
As the sole dark matter candidate, the lightest neutralino $\tilde\chi^0_1$ is required
to produce the correct thermal relic density. We require the neutralino relic density
to be in the $2\sigma$ range of the PLANCK and WMAP 9-year data,
$0.091 \leq \Omega h^2 \leq 0.138$, where a 10\% theoretical uncertainty
is included \cite{planck,wmap}.

\item[(iv)] {\em Muon g-2}.  we require NMSSM to explain the muon anomalous magnetic moment data
$\Delta a_\mu=(26.1\pm 8.0)\times 10^{-10}$ \cite{g-2} at $2\sigma$ level.

\item[(v)] {\em The absence of Landau pole}. We impose this constraint using $\textsf{NMSSMTools-4.0.0}$ \cite{nmssmtools}, where the interplay of $\lambda$ and $\kappa$ in the renormalization group running has been considered.

\item[(vi)] {\em LEP searches for SUSY}.
For the LEP experiments, the strongest constraints come from the chargino mass and the invisible $Z$ decay. We require $m_{\tilde{\chi}_1^\pm} \gtrsim 103 \, {\rm GeV}$ and the non-SM invisible decay width of $Z \to \tilde\chi^0_1\tilde\chi^0_1$ to be smaller than 1.71 MeV, which is consistent with the precision electroweak measurement result $\Gamma^{\rm non-SM}_{inv}<2.0$ MeV at $95\%$ confidence level \cite{zdecay}.

\item[(vii)] {\em Higgs data}.
Firstly, we consider the exclusion limits of the LEP, Tevatron and LHC in Higgs searches
with the package \textsf{HiggsBounds-4.0.0} \cite{higgsbounds}. This package also takes into account
the results of the LHC searches for non-SM Higgs bosons,
such as $H/A\to \tau^+\tau^-$ and $H^+\to \tau^+\nu_\tau$ \cite{htautau}.
Secondly, noticing that a light $h_1$ (or $a_1$) may induce the distinguished
signal $pp\to H \to h_1 h_1 (a_1 a_1) \to 4 \mu$, we consider the limitation of the $4\mu$ signal
on the parameter space using the latest CMS results \cite{4mu}.
Finally, since a large invisible branching ratio of the Higgs may be predicted in the light DM case,
we perform a global fit of the Higgs data using the package
\textsf{HiggsSignals-1.0.0} \cite{higgssignal}, where the systematics and correlations for the
signal rate predictions, luminosity and Higgs mass predictions are taken into account.
In our fit, we further consider the latest LHC results of Higgs invisible decay from the channel
$pp\to ZH$ \cite{zh_8}. We require our samples to be consistent with the Higgs data at $2\sigma$ level,
which corresponds to $\chi^2-\chi^2_{min}<4.0$ with $\chi^2$ obtained with the \textsf{HiggsSignals}
and $\chi^2_{min}$ denoting the minimum value of $\chi^2$ for the surviving samples in our scan.

\item[(viii)] {\em LHC searches for SUSY}.
Based on the 20 fb$^{-1}$ data collected at the 8 TeV run, the ATLAS and CMS collaborations performed a search for
the $\tilde\chi^0_2\tilde\chi^\pm_1$ production with $3\ell$+\met signal in a simplified model, where both $\tilde\chi_2^0$ and
$\tilde\chi_1^\pm$ are assumed to be wino-like with $Br(\tilde{\chi}_2^0 \to \tilde{\chi}_1^0 Z), Br(\tilde{\chi}_1^\pm \to
\tilde{\chi}_1^0 W^\pm) = 100\%$, and a 95\% C.L. upper limit on $\sigma \times BR$ was obtained on
the $m_{\tilde\chi^0_1}-m_{\tilde\chi^0_2}(m_{\tilde\chi^0_2} = m_{\tilde\chi^\pm_1}\simeq M_2)$ plane \cite{atlas_cms_ew}.

In this work, in order to implement this constraint we perform an analysis similar to \cite{Ellwanger_higgsino_singlino_sector}
with the code CheckMATE \cite{CheckMATE} for each sample surviving the constraints (i) - (vii). We consider the contributions from all
$\tilde\chi^0_i\tilde\chi^\pm_j$ ($i=2,3,4,5$ and $j=1,2$) associated production
processes to the signal, and calculate the production rates and the branching ratios with the code Prospino2 \cite{prospino}
and NMSDECAY \cite{NMSDECAY}, respectively. Our analysis indicates that this constraint can exclude effectively those samples with small values of $\mu$ below 115 GeV, and also some samples with moderate $\mu$ in the range from 115 GeV to 200 GeV. Nevertheless, compared to the results without considering this constraint, our conclusions do not change much such as the upper bounds on $Br(h_{SM} \to \tilde\chi^0_1 \tilde\chi^0_1, h_1 h_1, a_1 a_1)$ presented below .

\end{itemize}

In order to study the implication of the DM direct detection experiments on the NMSSM,
we also calculate the DM spin-independent elastic scattering cross section off nucleon with the formulae used in our previous work \cite{cao_cdms}. In getting the cross section, we set the parameter of the strange quark content in the nucleon as $f_{T_{s}} = 0.020$.

In the rest of this work, we categorize the DM by its component, i.e. either bino-like or singlino-like, in presenting our results. Since the interactions of the neutralinos with the Higgs bosons come from the following Lagrangian
\begin{eqnarray}
\cal{L} & = & \lambda ( s \tilde{H}_u^0 \tilde{H}_d^0 + h_u \tilde{H}_d^0 \tilde{S} + h_d \tilde{H}_u^0 \tilde{S} ) + \kappa s \tilde{S} \tilde{S} \nonumber \\
&& + \frac{i g_1}{\sqrt{2}} \tilde{B} (h_u \tilde{H}_u^0 - h_d \tilde{H}_d^0 ) -  \frac{i g_2}{\sqrt{2}} \tilde{W}^0 (h_u \tilde{H}_u^0 - h_d \tilde{H}_d^0 ),
\label{interaction}
\end{eqnarray}
where the fields $s$, $h_u$ and $h_d$ denote the neutral scalar parts of the Higgs superfields $\hat{S}$, $\hat{H}_u$ and $\hat{H}_d$, respectively, one can infer that if the DM is bino-like, the coupling strength of the $h_i \tilde{\chi}_1^0  \tilde{\chi}_1^0$ interaction is mainly determined by the higgsino-component in $\tilde{\chi}_1^0$, or more basically by the value of $\mu$. To be more specific, if $h_i$ is SM-like, the coupling strength is mainly determined by the first two terms in the second row of Eq.\ref{interaction}, while if $h_i$ is singlet-dominated, the coupling of $h_i \tilde{\chi}_1^0  \tilde{\chi}_1^0$ is mainly determined by the first term of Eq.\ref{interaction}. However, if the DM is singlino-like, the coupling strength is fundamentally determined by the parameters $\lambda$ and $\kappa$ and a low $\mu$ value may be helpful to enhance the coupling.

In this work, we are also interested in the couplings of the SM-like Higgs to light singlet-like scalars $h_1$ and $a_1$. These couplings are mainly determined by the following terms in the Higgs potential \cite{nmssmtools}
\begin{eqnarray}
V & = & \lambda^2 (|H_u|^2|S|^2 + |H_d|^2|S|^2) + \lambda\kappa (H_u \cdot H_dS^{*2} + \mathrm{h.c.}) \nonumber \\
&& + \kappa^2|S^2|^2 + (\lambda A_\lambda H_u \cdot
H_d S + \frac{1}{3} \kappa A_\kappa S^3 + \mathrm{h.c.}) + \cdots.
\label{Higgs_potential}
\end{eqnarray}
This equation indicates that, if $\lambda$ and $\kappa$ approach zero, the couplings $C_{h_{SM}h_1h_1},C_{h_{SM}a_1a_1}$
can not be very large; while if both of them have a moderate value, accidental cancelation is very essential to suppress the couplings.

\section{\label{results} Results and discussions}

\begin{figure}[t]
\includegraphics[width=13cm]{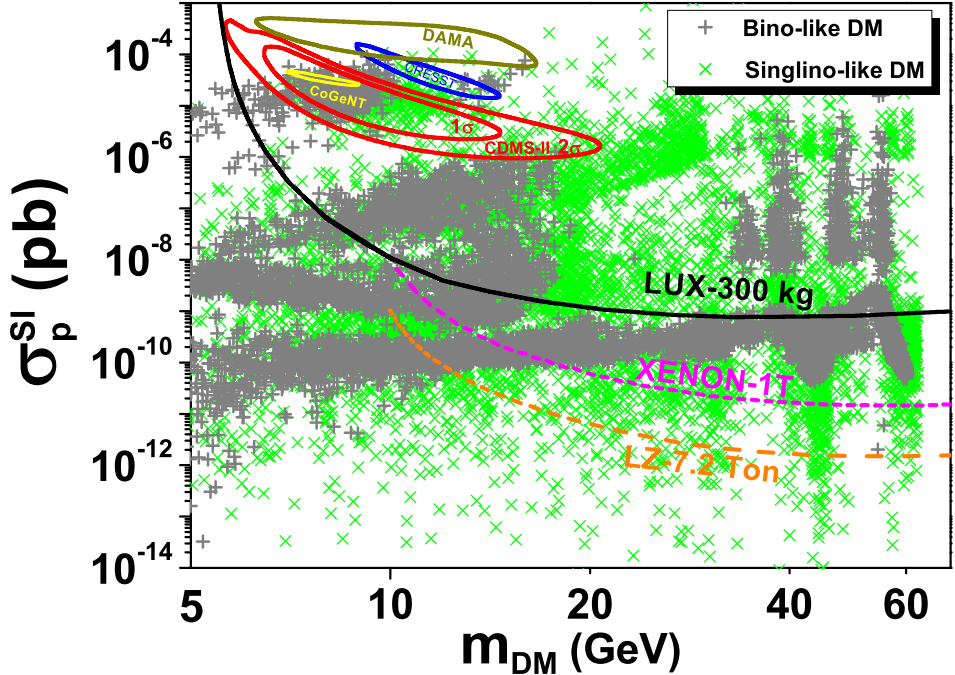}
\vspace*{-0.7cm}
\caption{The scatter plot of the NMSSM samples surviving various collider
experimental constraints and the dark matter relic density, projected on the plane of neutralino dark matter mass
versus spin-independent neutralino-nucleon scattering cross section.}
\label{fig1}
\end{figure}
In Fig.\ref{fig1} we project the samples surviving the above constraints
on the plane of neutralino dark matter mass
versus spin-independent neutralino-nucleon scattering cross section.
About this figure, we want to emphasize two points.
The first one is that some of the experimental constraints, such as the DM relic density and the Higgs data, play an important role
in limiting the parameter space of the NMSSM. So in the following, we pay special attention to investigate
how the samples in Fig.\ref{fig1} survive these constraints. The other one is that the various experimental constraints
will cut into the parameter space and the interplay among them is very complicated. As a result, the sample
distributions on the $m_{\tilde\chi^0_1}-\sigma^{\rm SI}_p$ plane might be very wired.
The strategy of analyzing this figure is to get a general picture of the current status of light DM confronting
the direct detection results and then focus on some interesting regions. As we will discuss later,
we will mainly focus on those samples that either can explain the CDMS-II results or can survive the first LUX
exclusion. We will not consider the up-right region ($m_{\tilde\chi^0_1}\gtrsim 20 \, {\rm GeV}, \sigma^{\rm SI}_p
\gtrsim 10^{-9} {\rm pb}$) in Fig.\ref{fig1} since it is not experimentally hinted.

After carefully analyzing our results, we have the following observations from Fig.\ref{fig1}:
\begin{enumerate}
  \item  In the NMSSM, DM as light as $5 \, {\rm GeV}$ is still allowed by the current Higgs data. Both the bino and singlino-like DM are capable of explaining the results of CDMS-II and CoGeNT, or surviving the current LUX results and future LUX exclusion limits.

  \item As pointed out in \cite{cao_cdms,dark-higgs}, light DM in the NMSSM may annihilate in the early universe through $s$-channel resonance effect of some mediators or into light Higgs scalar pair to get a correct relic density. We checked that, for $m_{\tilde{\chi}_1^0} \leq 35 \, {\rm GeV}$, singlino-like DM annihilated in the early universe mainly through the $s$-channel resonance effect of $h_1 (a_1)$ for the most case; while bino-like DM might annihilate either through the resonance effect or into $h_1$ ($a_1$) pair. We will discuss this issue in more detail later.

      In fact, the long thick band of grey samples (for bino-like DM) around $\sigma^{\rm SI}_p\sim 10^{-10} \, {\rm pb}$ exactly corresponds to the resonance case, and samples along this band are characterized by $m_{\tilde{\chi}^0_1} \simeq m_{\rm med}/2$ with $m_{\rm med}$ denoting the mediator mass. For $m_{\tilde\chi^0_1}\sim 45\, {\rm GeV}$ and $m_{\tilde\chi^0_1}\sim 60\, {\rm GeV}$, the mediator is $Z$ boson and the SM-like Higgs boson, respectively, while in other cases the mediator is either $h_1$ or $a_1$.  These conclusions can also apply to the singlino-like DM (see Fig.\ref{fig2}).

  \item For samples with $\sigma^{\rm SI}_p \gtrsim 10^{-9}\, {\rm pb}$, generally $h_1$ needs to be lighter than about $20 {\rm GeV}$ to push up the scattering rate. For the bino-like DM with mass varying from $17\, {\rm GeV}$ to $35\, {\rm GeV}$, such a light $h_1$ is difficult to obtain after considering the constraint from the relic density (see Fig.\ref{fig2}). While in the $Z$ ($h_{SM}$) resonance region, the relic density has rather weak limitation on $h_1$ properties. In this case, $h_1$ may be as light as several GeV so that the scattering rate is rather large, or the coupling $C_{h_1 \tilde\chi^0_1 \tilde\chi^0_1}$ may be greatly reduced to result in a relatively small $\sigma^{\rm SI}_p$.

  \item For bino-like DM, generally it is not easy to obtain samples with $\sigma^{\rm SI}_p \lesssim 10^{-11} \, {\rm pb}$. This is because the $h_{SM} \tilde\chi^0_1 \tilde\chi^0_1 $ interaction is still sizable even after considering the various constraints (see discussions on Fig.\ref{fig3}), and in this case, the $h_{SM}$-mediated contribution to the DM-nucleon scattering is important. However, in the extreme case when the bino-like DM is close to about $5\, {\rm GeV}$, due to the lower bound of $\mu$, the higgsino component in the DM will get further reduced and result in an even smaller $\sigma^{\rm SI}_p \sim 10^{-13} \, {\rm pb}$.

  \item When focusing on the XENON and LUX experiments, the bino-like and singlino-like DM exhibit quite different behaviors. The first LUX-300kg result can exclude a large part of the allowed parameter space, but still leaves both the bino-like and singlino-like light DM viable. The future XENON-1T and LUX-7.2Ton results can cut further deeply into the parameter space. Especially, they limit tightly the bino-like DM case and constrain most of the bino-like DM mass to be lower than about 17 GeV and 12 GeV, respectively, while the singlino-like DM can still survive leisurely.

  \item For bino-like DM samples there is a gap in the right half part of the CDMS-II $2\sigma$ region. This is due to the tension between the LHC Higgs data and the constraint from $\Upsilon \to h_1\gamma$. As discussed in \cite{french_dm} (and see Table I), the CDMS-II favored samples in bino-like DM scenario usually require a moderate $\lambda$ along with a moderate $\kappa$ to achieve the accidental cancelation in $C_{h_{SM}h_1h_1}, C_{h_{SM}a_1a_1}$ so that the SM-like Higgs decay to $h_1$ or $a_1$ pair is suppressed.
      While on the other hand, this may increase the effective coupling of $h_1$ to down-type fermions which is proportional to $\frac{\lambda_d m_{f_d}}{\sqrt{2}v} $ with $\lambda_d \approx \lambda \frac{v}{\mu} [ 1 + 2 (\frac{\mu}{m_Z})^2 ( \frac{A_\lambda}{\mu\tan\beta}-1  )  ]$ \cite{dark-higgs}, and receive constraint from the measurement of $\Upsilon \to h_1\gamma$.  We checked that most of the excluded bino-like DM samples in the gap have a relatively large $\lambda$, while the singlino-like DM samples usually correspond to a small $\lambda$ (see following discussion on Table I) and thus receive less constraint.

  \item Compared to bino-like DM which is restricted in certain areas on the $m_{\tilde\chi^0_1}-\sigma^{\rm SI}_p$ plane, singlino-like DM can spread nearly to the whole region of the plane. This reflects the fact that singlino-like DM is more adaptable in light DM physics.
\end{enumerate}

\begin{table}
\centering
\vspace{0.3cm}
\begin{tabular}{| c || l | l || l | l |}
\hline
 & \multicolumn{2}{|c||}{bino-like} & \multicolumn{2}{|c|}{singlino-like} \\
\cline{2-5}
  & \multicolumn{1}{|c|}{CDMS-II} & \multicolumn{1}{|c||}{LUX} & \multicolumn{1}{|c|}{CDMS-II} & \multicolumn{1}{|c|}{LUX} \\
\hline
$M_1$ & (8\ ,\ 22)  & (4\ ,\ 39)  & (-600\ ,\ -110) & (-600\ ,\ -30) \\
\hline
$M_2$ & (320\ ,\ 600) & (320\ ,\ 600)  & (320\ ,\ 600) & (320\ ,\ 600) \\
\hline
$\mu$ & (160\ ,\ 225) & (157\ ,\ 450)  & (115\ ,\ 220) & (119\ ,\ 480) \\
\hline
$\tan\beta$ & (14\ ,\ 28) & (6\ ,\ 40)  & (7\ ,\ 29) & (7\ ,\ 37) \\
\hline
$\lambda$ & (0.28\ ,\ 0.49) & (0.015\ ,\ 0.59)  & (0.08\ ,\ 0.25) & (0.06\ ,\ 0.3) \\
\hline
$\kappa$ & (0.29\ ,\ 0.57) & (0\ ,\ 0.6)  & (-0.01\ ,\ 0.02) & (-0.03\ ,\ 0.02) \\
\hline
$A_\lambda$ & (2400\ ,\ 4800) & (1050\ ,\ 5000) & (1070\ ,\ 4990) & (1200\ ,\ 5000) \\
\hline
$A_\kappa$ & (-1100\ ,\ -630) & (-1300\ ,\ 0)  & (-80\ ,\ 60) & (-120\ ,\ 110) \\
\hline
\end{tabular}
\caption{The ranges of relevant NMSSM input parameters corresponding to part of the samples in Fig.1,
which predict a DM lighter than 35 GeV and meanwhile can explain the CDMS-II at 2$\sigma$
level or survive the LUX-300kg exclusion limit. Parameters with the mass dimension are in the unit of GeV.}
\label{table1}
\end{table}

In the following, we concentrate on the samples in Fig.1 that can either explain the CDMS-II experiment at $2\sigma$ level or survive the LUX-300kg exclusion limit. Since the results of the CDMS-II and LUX experiments are so incompatible, it would be interesting to investigate the difference of these two types of samples. To simply our analysis, we mainly consider the samples predicting a DM lighter than about 35 GeV. These samples are not easy to obtain with traditional random scan method when exploring the SUSY parameter space due to its rather specific particle spectrum, but as we will see below, the underlying physics of these samples are clear and easy to understood. In Table \ref{table1}, we list the ranges of relevant NMSSM input parameters corresponding to these samples, which are classified by the component of the DM (i.e. bino-like or singlino-like) and meanwhile by its scattering cross section off the nucleon (i.e. can explain the CDMS-II results at $2\sigma$ level or survive the LUX-300kg exclusion limit).

From Table I, one can learn the following facts:
\begin{itemize}

\item The survived parameter ranges for LUX-safe samples are generally wider than those of CDMS-II preferred samples. This is totally expectable from the experimental data of LUX and CDMS-II. On the $m_{\tilde\chi^0_1}-\sigma^{\rm SI}_p$ plane, CDMS-II $2\sigma$ region is constrained in a relatively narrow range $5.7 \, {\rm GeV} \lesssim m_{\tilde\chi^0_1} \lesssim 20.7 \, {\rm GeV}$ and $10^{-6} {\rm \, pb} \lesssim \sigma^{\rm SI}_p \lesssim 10^{-4} {\rm \, pb}$. To survive the first LUX exclusion, however, a properly large $m_{h_1}$ for a certain $m_{\tilde\chi^0_1}$ will be enough. $m_{\tilde\chi^0_1}$ can cover the whole range $(5 \, {\rm GeV} , 60 \, {\rm GeV})$ and $\sigma^{\rm SI}_p$ can vary from $10^{-13} {\rm \, pb}$ to $10^{-3} {\rm \, pb}$. Therefore, compared to CDMS-II region, there will be more freedom for the parameter space to satisfy the LUX exclusion.

\item To obtain a DM lighter than 35 GeV, one needs to have $|M_1| \lesssim 40 \, {\rm GeV}$ for the bino-like DM and $|\kappa| \ll \lambda$ for the singlino-like DM. This can be easily understood from the neutralino mass matrix \cite{nmssm_review}.
\begin{eqnarray}
{\cal M}_0 =
\left( \begin{array}{ccccc}
M_1 & 0 & -\frac{g_1 v_d}{\sqrt{2}} & \frac{g_1 v_u}{\sqrt{2}} & 0 \\
& M_2 & \frac{g_2 v_d}{\sqrt{2}} & -\frac{g_2 v_u}{\sqrt{2}} & 0 \\
& & 0 & -\mu & -\lambda v_u \\
& & & 0 & -\lambda v_d \\
& & & & \frac{2\kappa}{\lambda} \mu
\end{array} \right),
\label{neutralino_mass_matrix}
\end{eqnarray}
where $g_1$ and $g_2$ are gauge couplings, and $v_u$ and $v_d$ are Higgs vacuum expectation values.
In fact, a simple estimation can be made for singlino-like DM mass. Table \ref{table1} shows that $|\kappa|$ is usually at least one order smaller than $\lambda$. Assuming $|\kappa|/\lambda\sim 1/20$ and $\mu\sim 200\, {\rm GeV}$, we will have $m_{\tilde\chi^0_1}\sim 20 \, {\rm GeV}$.

\item The CDMS-II samples usually have $|\mu| \lesssim 225 \, {\rm GeV}$ for both bino-like and singlino-like DM. The underlying reason is that a small value of $\mu$ and consequently a sufficient amount of higgsino component in the DM is helpful to increase the coupling strength of the DM with the light Higgs bosons. This will in return push up the rate of the DM-nucleon scattering which is required by the CDMS-II results.

\item More interestingly, we find that for samples in the whole range of $m_{\tilde{\chi}_1^0} \lesssim 35\, {\rm GeV}$, the value of $\mu$ is upper bounded by about $450 \, {\rm GeV}$ and $480 \, {\rm GeV}$ for bino-like and singlino-like DM, respectively. Two reasons can account for this. The first one is that in our scan, we required the NMSSM to explain the muon anomalous magnetic moment. The parameter $\mu$ influences the contribution of the NMSSM to the moment through chargino and neutralino mass, and a large value of $\mu$ will reduce the contribution significantly. Another important reason is that, as mentioned above and also discussed below, in order to get a correct DM relic density, a light $h_1$ or $a_1$ must be present. Noting that $\mu$ enters explicitly the squared mass of the singlet scalar \cite{nmssm_review}, one can infer that too large values of $\mu$ can not be favored to get the desired light scalar masses.

    We also want to emphasize that, for the bino-like DM, an upper bound of $\mu$ will result in a lower limit of the higgsino component in the DM and thus a lower bound of the invisible branching ratio for $h_{SM} \to \tilde{\chi}_1^0 \tilde{\chi}_1^0$. This can be explicitly seen in the left panel of Fig.\ref{fig3} below.

\item For singlino-like DM case, both $\lambda$ and $\kappa$ are small and especially, $|\kappa|$ is very close to 0. As indicated by Eqs.(\ref{interaction},\ref{Higgs_potential},\ref{neutralino_mass_matrix}), the couplings of SM-like Higgs boson to DM and also to the light Higgs scalars $h_1,a_1$ will usually be suppressed. This can result in a $\sigma^{\rm SI}_p$ as low as  $10^{-14}\,{\rm pb}$ (see Fig.\ref{fig1}) and also a relatively small rate for the decays $h \to \chi_1^0 \chi_1^0, h_1 h_1, a_1 a_1$ (see Fig.\ref{fig3} and Fig.\ref{fig4}). While for the bino-like DM case with a moderate value of $\lambda$ and $\kappa$, accidental cancelation is very essential to suppress the couplings of $h_{SM}$ to $\chi_1^0 \chi_1^0, h_1 h_1, a_1 a_1$ and obtain an allowed Higgs signal.
\end{itemize}

As discussed in Fig.\ref{fig1}, given $m_{\tilde{\chi}_1^0} \lesssim 35 \, {\rm GeV}$, at least one light scalar is needed to accelerate the annihilation.
In order to illustrate this feature, in Fig.\ref{fig2} we project the $m_{\tilde{\chi}_1^0} \lesssim 35 \, {\rm GeV}$ samples of Fig.1 which can explain the CDMS-II results at 2$\sigma$
level or survive the LUX-300kg exclusion limits on the plane of DM mass versus ${\rm min}(m_{h_1},m_{a_1})$. Red codes represent samples suggested by the CDMS-II experiment and meanwhile satisfying $m_{h_1}<m_{a_1}$, while cyan (blue) codes correspond to samples surviving the LUX-300kg exclusion limits and also satisfying $m_{h_1}<m_{a_1}$ ($m_{h_1}>m_{a_1}$). Note that due to the large scattering cross section favored by the CDMS-II results, a light $h_1$ is needed (as the t-channel propagator) and the case $m_{h_1} > m_{a_1}$ is absent. From Fig.\ref{fig2} we have the following observations:
\begin{figure}[tl]
\centering\includegraphics[width=16cm]{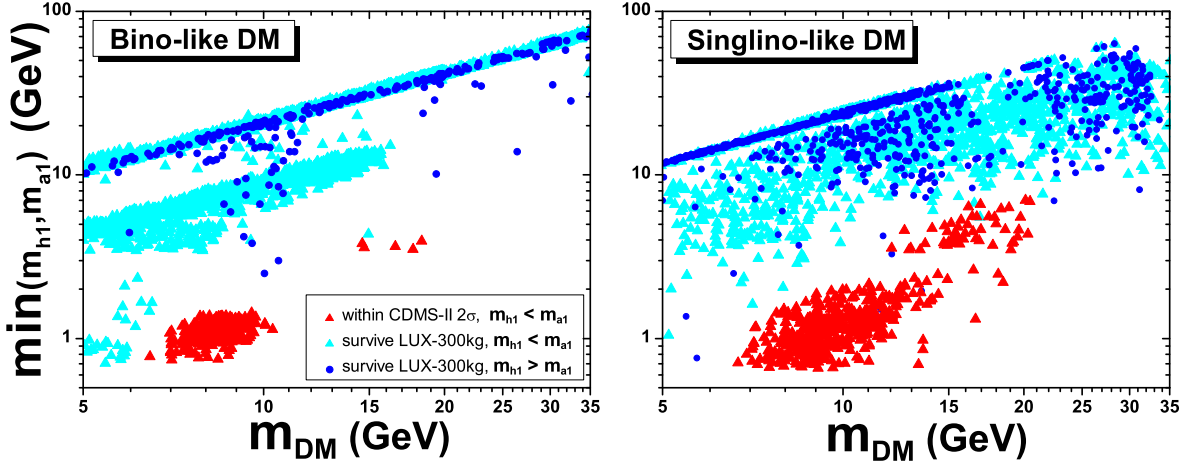}
\vspace*{-0.7cm}
\caption{Scatter plot of the $m_{\tilde{\chi}_1^0} \lesssim 35 \, {\rm GeV}$ samples in Fig.\ref{fig1} which can explain the CDMS-II results at 2$\sigma$ level or survive the LUX-300kg exclusion limits, projected on the plane of DM mass versus ${\rm min}(m_{h_1},m_{a_1})$. Red codes represent samples suggested by the CDMS-II experiment and meanwhile satisfying $m_{h_1}<m_{a_1}$, while cyan(blue) codes correspond to samples surviving the LUX-300kg exclusion limits and also satisfying $m_{h_1}<m_{a_1}$ ($m_{h_1}>m_{a_1}$). Note that due to the large scattering cross section favored by the CDMS-II results, a light $h_1$ is needed (as the t-channel propagator) and the case $m_{h_1} > m_{a_1}$ is absent.}
\label{fig2}
\end{figure}
\begin{enumerate}
  \item  In both bino-like and singlino-like DM scenario, the straight line ${\rm min}(m_{h_1},m_{a_1})\sim 2m_{\tilde{\chi}_1^0}$ is very obvious, which corresponds to the s-channel resonance effect of $h_1$ or $a_1$. However, in the singlino-like scenario with $m_{\tilde{\chi}_1^0} \gtrsim 18 \, {\rm GeV}$, there are some small regions where the line seems to be not continuous. In fact, this is not the case. We checked that there still exits a scalar (either $h_1$ or $a_1$) with mass around $2 m_{\tilde{\chi}_1^0}$. It is just that this scalar does not correspond to the lightest Higgs boson. Moreover, for the scalars shown in Fig.\ref{fig2}, we checked that they are highly singlet-dominated, which agree with previous study in \cite{cao_light_higgs}.

  \item  Since $h_1$ contributes to the spin-independent DM-nucleon scattering as the t-channel propagator \cite{cao_cdms}, a very light $h_1$ is needed to explain the CDMS-II result. For the bino-like DM, the CDMS-II samples are mainly distributed in low $m_{\tilde{\chi}_1^0}$ region with $m_{h_1}$ upper bounded by about 4 GeV, while for the singlino-like DM, the corresponding samples spread a larger region in $m_{\tilde{\chi}_1^0}-m_{h_1}$ plane. Moreover, when focusing on the CDMS-II samples, we checked that if the DM is bino-like, the channel $\chi_1^0 \chi_1^0 \to h_1 h_1$ plays the dominant role in contributing to the DM annihilation, while if the DM is singlino-like,
      the s-channel resonance effect is the main contribution.

  \item Since the constraint from the LUX-300kg data on the scattering rate is rather weak in the very light DM region, $h_1$ as light as 1 GeV is still allowed for  $m_{\tilde{\chi}_1^0} \lesssim 7 \, {\rm GeV}$. With the increase of DM mass, the constraint becomes much stronger and $h_1$ generally needs to be heavier than about 10 GeV for $m_{\tilde{\chi}_1^0} \gtrsim 25 \, {\rm GeV}$ in both scenarios.

\end{enumerate}

\begin{figure}[t]
\includegraphics[width=16cm]{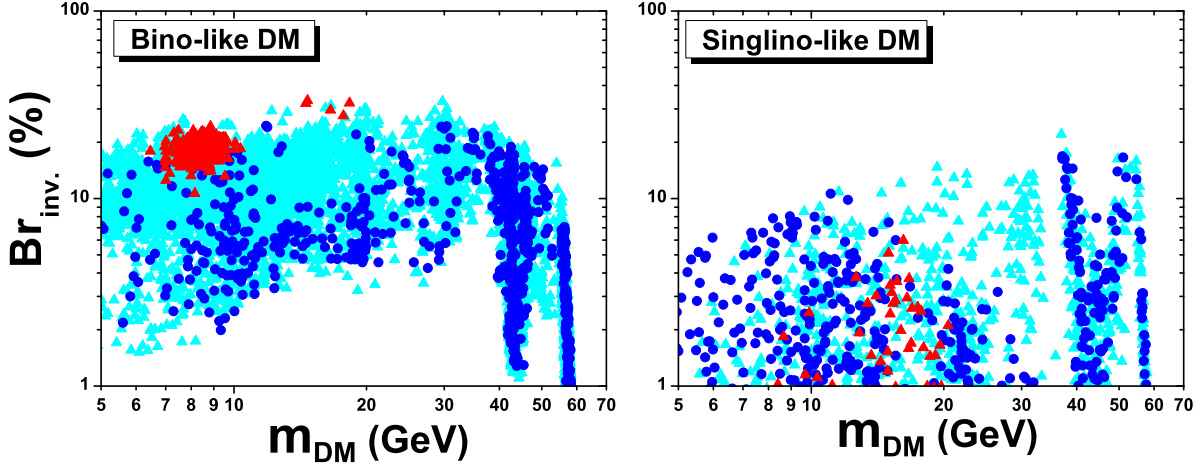}
\vspace*{-0.7cm}
\caption{Same as Fig.\ref{fig2}, but projected on the plane of the invisible branching fractions of the SM-like Higgs boson versus DM mass, and extended the DM mass to about 60 GeV.}
\label{fig3}
\end{figure}

For the SM-like Higgs boson, since the decay channel $h_{SM} \to \tilde{\chi}_1^0  \tilde{\chi}_1^0$ is opened when $m_{\tilde{\chi}_1^0} < m_{h_{SM}}/2$, one can expect that the Higgs data will impose rather tight constraints on this decay rate.
In Fig.\ref{fig3}, we show the samples of Fig.\ref{fig2} on the plane of $Br(h_{SM} \to \tilde\chi^0_1 \tilde\chi^0_1)$ versus DM mass and extend the DM mass to about 60 GeV. We have the following observations:
\begin{enumerate}
  \item The current Higgs data still allow for an invisible decay branching ratio as large as $30\%$ at $2\sigma$ level. The tolerance of  such a large invisible branching ratio is owe to the large uncertainties of the current data, especially the fact that ATLAS and CMS data point to two opposite directions in the di-photon rate. Obviously, an invisible decay branching ratio reaching $30\%$ may be easily tested at the 14 TeV LHC with $\mathcal{L}=100$ fb$^{-1}$, where a 95\% C.L. upper limit on the invisible decay, i.e. $Br_{\rm inv}<18\%$, can be imposed \cite{zh_8,zh_14}.
  \item In the bino-like DM scenario, due to the necessary higgsino component in the DM required by an efficient DM annihilation rate, the interaction between DM and $h_{SM}$ can be relatively large. As a result, $Br(h_{SM} \to \tilde\chi^0_1 \tilde\chi^0_1)$ as large as $30\%$ is possible. Note that for the CDMS-II samples, $Br(h_{SM} \to \tilde\chi^0_1 \tilde\chi^0_1)$ is always larger than about $10\%$. The underlying reason is that, as we mentioned earlier, the channel $\chi_1^0 \chi_1^0 \to h_1 h_1$ plays an important role in contributing to the DM annihilation. This requires the strength of the $h_1 \tilde{\chi}_1^0 \tilde{\chi}_1^0$ interaction to be sufficiently large, and so is the  $h_{SM} \tilde{\chi}_1^0 \tilde{\chi}_1^0$ interaction. Also note that since $\mu$ is upper bounded for $m_{\tilde{\chi}^0_1} \lesssim 35 {\rm GeV}$ (see Table \ref{table1}), generally there is a lower bound of $Br(h_{SM} \to \tilde{\chi}^0_1 \tilde{\chi}^0_1)$.

  \item In the singlino-like scenario, since the $h_{SM} {\tilde{\chi}_1^0}{\tilde{\chi}_1^0}$ coupling is determined by $\lambda$ and $\kappa$ and Table \ref{table1} indicates that these two parameters are generally small, $Br(h_{SM} \to \tilde\chi^0_1 \tilde\chi^0_1)$ is usually suppressed and can reach about $20\%$ in the optimal case.
\end{enumerate}

\begin{figure}[t]
\includegraphics[width=16cm]{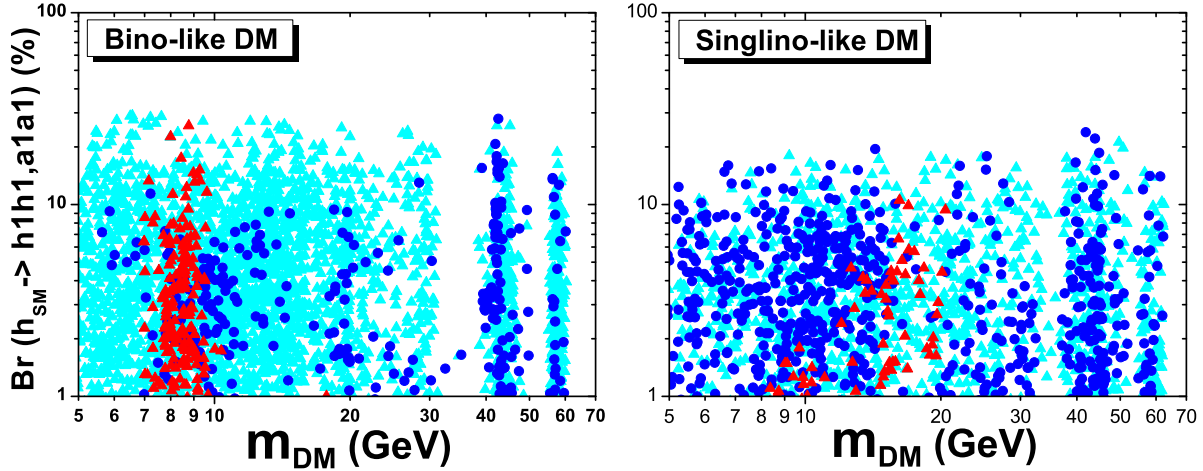}
\vspace*{-0.7cm}
\caption{Same as Fig.\ref{fig3}, but showing the branching fraction of decays $h_{SM} \to h_1 h_1,a_1a_1$ versus DM mass.}
\label{fig4}
\end{figure}

Due to the existence of light scalars in light DM scenario, the SM-like Higgs may also decay into the lighter scalars, $h_{SM} \to h_1 h_1({\rm or}~ a_1a_1)$. Unlike the $h_{SM}\tilde\chi^0_1\tilde\chi^0_1$ coupling, the coupling strengthes of $h_{SM}$ to these scalars are mainly determined by $\lambda$ and $\kappa$ (see Eq.(\ref{Higgs_potential}) and also note that both $h_1$ and $a_1$ are highly singlet-dominated \cite{cao_light_higgs}). Consequently, according to Table \ref{table1}, the maximum decay rate in the bino-like DM scenario should in principle be larger than that in the singlino-like case. Similar to Fig.\ref{fig3}, we show the total branching fractions of these two decays versus DM mass in Fig.\ref{fig4}. One can learn that this branching ratio can reach $30\%$ in the bino-like DM scenario, while in the singlino-like case the maximum can only reach about $20\%$.

\section{\label{conclusion}Conclusion}
Under current experimental constraints including the latest LHC Higgs data and the
dark matter
relic density, we examined the status of a light NMSSM dark matter and confronted it
with the direct detection results of CoGeNT, CDMS-II and LUX. We have the
following observations: (i) A dark matter as light as 8 GeV is still
allowed and its scattering cross section off the nucleon can be large enough to
explain the CoGeNT/CDMS-II favored region;
(ii) The LUX data can exclude a sizable part of the allowed parameter space, but
still leaves a light dark matter viable;
(iii) The SM-like Higgs boson can decay into the light dark matter pair and
its branching ratio can reach $30\%$ at $2\sigma$ level under the current LHC
Higgs data, which may be covered largely at the 14 TeV LHC experiment.

\vspace*{1.0cm}
We thank Nima Arkani-Hamed, Archil Kobakhidze, Yang Zhang and Jie Ren for helpful discussions.
This work was supported by the ARC Center of Excellence for Particle Physics at the Tera-scale,
by the National Natural Science Foundation of China (NNSFC)
under grant No. 10821504, 11222548, 11305049 and 11135003, and also by Program for New Century Excellent
Talents in University.


\begin{thebibliography}{99}

\bibitem{DM_before_1}
  D.~A.~Vasquez, G.~Belanger, C.~Boehm, A.~Pukhov and J.~Silk,
  Phys.\ Rev.\ D {\bf 82}, 115027 (2010)
  [arXiv:1009.4380 [hep-ph]].

\bibitem{DM_before_2}
  D.~Albornoz Vasquez, G.~Belanger and C.~Boehm,
  Phys.\ Rev.\ D {\bf 84}, 095008 (2011)
  [arXiv:1107.1614 [hep-ph]].

\bibitem{DM_before_3}
  G.~Belanger, G.~Drieu La Rochelle, B.~Dumont, R.~M.~Godbole, S.~Kraml and S.~Kulkarni,
  Phys.\ Lett.\ B {\bf 726}, 773 (2013)
  [arXiv:1308.3735 [hep-ph]].

\bibitem{DM_before_4}
  K.~Schmidt-Hoberg, F.~Staub and M.~W.~Winkler,
  Phys.\ Lett.\ B {\bf 727}, 506 (2013)
  [arXiv:1310.6752 [hep-ph]].

\bibitem{DM_before_5}
  K.~-Y.~Choi and O.~Seto,
  Phys.\ Rev.\ D {\bf 88}, 035005 (2013)
  [arXiv:1305.4322 [hep-ph]].

\bibitem{DM_before_6}
  C.~Boehm, P.~S.~B.~Dev, A.~Mazumdar and E.~Pukartas,
  JHEP {\bf 1306}, 113 (2013)
  [arXiv:1303.5386 [hep-ph]].

\bibitem{DM_before_7}
  B.~Kyae and J.~-C.~Park,
  arXiv:1310.2284 [hep-ph].

\bibitem{CDMS2}
  R.~Agnese {\it et al.} [CDMS Collaboration],
  arXiv:1304.4279 [hep-ex].

\bibitem{CoGeNT1}
  C.~E.~Aalseth {\it et al.}  [CoGeNT Collaboration],
  Phys.\ Rev.\ Lett.\  {\bf 106}, 131301 (2011).

\bibitem{CoGeNT2}
  C.~E.~Aalseth {\it et al.} [CoGeNT Collaboration],
  Phys.\ Rev.\ Lett.\  {\bf 107}, 141301 (2011).

\bibitem{CRESST}
  G.~Angloher  {\it et al.} [CRESST Collaboration],
  Eur.\ Phys.\ J.\ C {\bf 72}, 1971 (2012).

\bibitem{DAMA}
  R.~Bernabei {\it et al.}  [DAMA Collaboration],
  Eur.\ Phys.\ J.\ C {\bf 56}, 333 (2008).

\bibitem{Fermi_LAT_experiment}
  M.~Ackermann {\it et al.}  [LAT Collaboration],
  Phys.\ Rev.\ D {\bf 86}, 022002 (2012)
  [arXiv:1205.2739 [astro-ph.HE]].

\bibitem{Fermi_LAT_analysis_1}
  L.~Goodenough and D.~Hooper,
  arXiv:0910.2998 [hep-ph].

\bibitem{Fermi_LAT_analysis_2}
  D.~Hooper and L.~Goodenough,
  Phys.\ Lett.\ B {\bf 697}, 412 (2011)
  [arXiv:1010.2752 [hep-ph]].

\bibitem{Fermi_LAT_analysis_3}
  D.~Hooper and T.~Linden,
  Phys.\ Rev.\ D {\bf 84}, 123005 (2011)
  [arXiv:1110.0006 [astro-ph.HE]].

\bibitem{Fermi_LAT_analysis_4}
  K.~N.~Abazajian and M.~Kaplinghat,
  Phys.\ Rev.\ D {\bf 86}, 083511 (2012)
  [arXiv:1207.6047 [astro-ph.HE]].

\bibitem{Fermi_LAT_analysis_5}
  W.~-C.~Huang, A.~Urbano and W.~Xue,
  arXiv:1307.6862 [hep-ph].

\bibitem{XENON100}
  E.~Aprile {\it et al.}  [XENON100 Collaboration],
  Phys.\ Rev.\ Lett.\  {\bf 109}, 181301 (2012).

\bibitem{lux}
  D.~S.~Akerib {\it et al.}  [LUX Collaboration],
  arXiv:1310.8214 [astro-ph.CO].

\bibitem{CDEX_1}
  k.~J.~Kang {\it et al.}  [CDEX Collaboration],
  arXiv:1303.0601 [hep-ph].

\bibitem{CDEX_2}
  J.~Su {\it et al.}  [CDEX Collaboration],
  arXiv:1402.4591 [nucl-ex].

\bibitem{cao_cdms}
  J.~Cao, K.~-i.~Hikasa, W.~Wang, J.~M.~Yang and L.~-X.~Yu,
  JHEP {\bf 1007}, 044 (2010)
  [arXiv:1005.0761 [hep-ph]].

\bibitem{dark-higgs}
  P.~Draper, T.~Liu, C.~E.~M.~Wagner, L.~-T.~Wang and H.~Zhang,
  Phys.\ Rev.\ Lett.\  {\bf 106}, 121805 (2011)
  [arXiv:1009.3963 [hep-ph]].

\bibitem{nmssm_review}
  U.~Ellwanger, C.~Hugonie and A.~M.~Teixeira,
  Phys.\ Rept.\  {\bf 496}, 1 (2010).

\bibitem{MSSM_DM_han}
  T.~Han, Z.~Liu and A.~Natarajan,
  JHEP {\bf 1311}, 008 (2013)
  [arXiv:1303.3040 [hep-ph]].

\bibitem{MSSM-light-DM-sbottom}
  A.~Arbey, M.~Battaglia and F.~Mahmoudi,
  arXiv:1308.2153 [hep-ph].

\bibitem{cao_light_higgs}
  J.~Cao, F.~Ding, C.~Han, J.~M.~Yang and J.~Zhu,
  JHEP {\bf 1311}, 018 (2013)
  [arXiv:1309.4939 [hep-ph]].

\bibitem{belanger_fit}
  G.~Belanger, B.~Dumont, U.~Ellwanger, J.~F.~Gunion and S.~Kraml,
  Phys.\ Rev.\ D {\bf 88}, 075008 (2013)
  [arXiv:1306.2941 [hep-ph]].

\bibitem{lhc_data}
  CMS Collaboration,
  CMS-PAS-HIG-13-005;
  ATLAS Collaboration,
  ATLAS-CONF-2013-034.

\bibitem{atlas_cms_ew}
  ATLAS Collaboration,
  ATLAS-CONF-2013-035;
  CMS Collaboration,
  CMS-PAS-SUS-13-006.

\bibitem{Ellwanger_higgsino_singlino_sector}
  U.~Ellwanger,
  JHEP {\bf 1311}, 108 (2013)
  [arXiv:1309.1665 [hep-ph]].

\bibitem{french_dm}
  J.~Kozaczuk and S.~Profumo,
  arXiv:1308.5705 [hep-ph].

\bibitem{higgssignal}
  P.~Bechtle, S.~Heinemeyer, O.~Stal, T.~Stefaniak and G.~Weiglein,
  arXiv:1305.1933 [hep-ph].

\bibitem{zh_8}
  ATLAS Collaboration,
  ATLAS-CONF-2013-011.

\bibitem{zdecay}
  S.~Schael {\it et al.}  [ALEPH and DELPHI and L3 and OPAL and SLD and LEP Electroweak Working Group and SLD Electroweak Group and SLD Heavy Flavour Group Collaborations],
  Phys.\ Rept.\  {\bf 427}, 257 (2006)
  [hep-ex/0509008].

\bibitem{nmssmtools}
  U. Ellwanger, J. F. Gunion and C. Hugonie, JHEP {\bf 0502}, 066 (2005);
  U. Ellwanger and C. Hugonie, Comput. Phys. Commun. {\bf 175}, 290 (2006);
  G. Degrassi {\it et al.}, Eur. Phys. J. C {\bf 28} (2003) 133.

\bibitem{upsilon}
   B.~Aubert {\it et al.}  [BaBar Collaboration],
  Phys.\ Rev.\ Lett.\  {\bf 103}, 081803 (2009)
  [arXiv:0905.4539 [hep-ex]].

\bibitem{bsg}
  Y.~Amhis {\it et al.}  [Heavy Flavor Averaging Group Collaboration],
  arXiv:1207.1158 [hep-ex].

\bibitem{bsmm}
  Raij {\it et al.}  [LHCb Collaboration],
  Phys.\ Rev.\ Lett.\  {\bf 110}, 021801 (2013).

\bibitem{planck}
  P.~A.~R.~Ade {\it et al.}  [Planck Collaboration],
  arXiv:1303.5076 [astro-ph.CO].

\bibitem{wmap}  
  G.~Hinshaw {\it et al.}  [WMAP Collaboration],
  Astrophys.\ J.\ Suppl.\  {\bf 208}, 19 (2013).

\bibitem{g-2}
  K.~Hagiwara  {\it et al.},
  J.\ Phys.\ G {\bf 38}, 085003 (2011).

\bibitem{higgsbounds}
  P.~Bechtle {\it et al.},
  Comput.\ Phys.\ Commun.\  {\bf 181}, 138 (2010);
  Comput.\ Phys.\ Commun.\  {\bf 182}, 2605 (2011);
  PoS CHARGED {\bf 2012}, 024 (2012)  [arXiv:1301.2345 [hep-ph]].

\bibitem{htautau}
  CMS Collaboration,
  CMS-PAS-HIG-12-050;
  ATLAS Collaboration,
  ATLAS-CONF-2012-011.

\bibitem{4mu}
  CMS Collaboration,
  CMS-PAS-HIG-13-010.

\bibitem{CheckMATE}
  M.~Drees, H.~Dreiner, D.~Schmeier, J.~Tattersall and J.~S.~Kim,
  arXiv:1312.2591 [hep-ph].

\bibitem{prospino}
  W.~Beenakker, M.~Klasen, M.~Kramer, T.~Plehn, M.~Spira and P.~M.~Zerwas,
  Phys.\ Rev.\ Lett.\  {\bf 83}, 3780 (1999)  [Erratum-ibid.\  {\bf 100}, 029901 (2008)]  [hep-ph/9906298].

\bibitem{NMSDECAY}
  D.~Das, U.~Ellwanger and A.~M.~Teixeira,
  Comput.\ Phys.\ Commun.\  {\bf 183}, 774 (2012)
  [arXiv:1106.5633 [hep-ph]].

\bibitem{zh_14}
  D.~Ghosh,  {\it et al.},
  Phys.\  Lett.\ B {\bf  725}, 344 (2013);
  CMS-PAS-HIG-13-013,
  CMS-PAS-HIG-13-018,
  CMS-PAS-HIG-13-028,

\end{thebibliography}
\end{document}